\documentclass[aip,graphicx,rsi,reprint]{revtex4-2}
\usepackage{tikz,siunitx}
\usepackage{amsmath}
\usepackage{amssymb}
\usepackage{bbold}
\usepackage{bm}
\usepackage{listings}
\usepackage{mathtools}
\usetikzlibrary{angles,positioning}

\usepackage{hyperref}
\usepackage[capitalise,noabbrev]{cleveref}

\NewDocumentCommand{\codeword}{v}{%
\texttt{\textcolor{black}{#1}}%
}

\draft 

\begin{document}


\title[]{Grazing incidence X-ray scattering alignment using the area detector} 



\author{E. Tortorici}
\email[]{edward.tortorici@colorado.edu}
\affiliation{ 
Department of Physics, University of Colorado Boulder, 80309-0215, United States
}

\author{C. T. Rogers}%
 \homepage{https://spot.colorado.edu/~rogersct/}
 \affiliation{ 
Department of Physics, University of Colorado Boulder, 80309-0215, United States
}

\date{\today}

\begin{abstract}
    Grazing incidence X-ray scattering experiments are designed to achieve strong scattering signals from materials, such as molecular monolayers, island films, or thin films that are localized to the surfaces of flat substrates. Optimal signals can be achieved with precise alignment of a substrate surface with the X-ray beam. Here, we outline a simple method that utilizes the area detector, generally available on such systems, to observe reflections from the sample to determine the sample-detector distance and the motor positions corresponding to the film being parallel to and centered in the beam. Observations of the reflected and transmitted beams are used to determine the critical angle of the sample and inform ideal motor angles that will lead to scattered X-ray intensity enhancement.
\end{abstract}


\maketitle 



\section{\label{sec:intro}Introduction}


    Grazing incidence X-ray scattering (GIXS) experiments allow one to probe the structure and texture of materials localized to the flat surfaces of substrates.\cite{film-texture-sim,giwaxs-perovskite,giwaxs-principles,giwaxs-tool,quantify-film-alignment} Molecular monolayers, island films, thin films, and other surface-localized materials lack bulk, and only a small amount of the beam will typically intersect the material localized to a substrate. A grazing-angle of incidence can lead to scattering enhancements,\cite{enhanced-xrays,waveguide-enhanced-gixs} and to achieve optimal scattering intensity, GIXS experiments require precise alignment of the sample surface relative to the beam. Alignment methods often vary, given the variety in the design of experimental systems.\cite{giwaxs-align} In this paper, we propose a simple, fast, and reliable methodology for aligning a GIXS experiment that utilizes the area detector, already available to a system used for grazing incidence wide-angle X-ray scattering and grazing incidence small-angle X-ray scattering (GIWAXS and GISAXS respectively). The area detector is used to spatially resolve (1) specular reflections in order to accurately determine the grazing-incidence angle and the sample-detector distance and (2) transmitted X-rays in order to determine the refractive index of (and therefore the critical angle for) the material. Even if the sample stage is itself well-machined, aligned, and characterized, alignment is required for each individual experiment because substrates often do not have parallel top and bottom surfaces and unless a clean vacuum chuck is used, the bottom of the surface of the substrate may not be parallel to the stage.

\section{Example Application}

\begin{figure}[h!]
\centering
\includegraphics{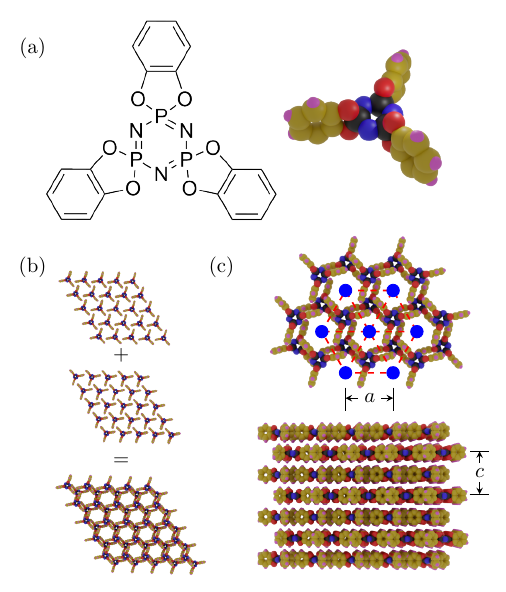}
\caption[TPP molecule and hexagonal crystal phase]{
    (a) TPP is an organic molecule with a posphazene ring oxygen-bonded to three phenyl rings. (b) TPP forms a hexagonal bilayer. (c) TPP crystals are flexible, and the $a$ lattice parameter can vary up to \SI{10}{\percent} to accommodate guests (labeled as blue circles).}
\label{fig:tpp-crystal}
\end{figure}

    As an example, we use GIXS to study organic thin-film inclusion compounds using hexagonal tris(\emph{o}-phenylene-dioxy)cyclotriphosphazene (TPP) host crystals. Organic materials are composed of low-$Z$ atoms, and are therefore poor scatterers due to having low electron densities, leading to even lower scattering intensities compared to materials containing heavier atoms.\cite{cullity3rd} TPP is known to form hexagonal crystals with inclusion ducts that can trap guest molecules to form inclusion compounds\cite{allcock-inclusion-1, allcock-inclusion-nmr, allcock-molecular-motion, allcock-monoclinic-transition, allcock-synthesis-2000, allcock-xray-1973, bulk-inclusion-2013, arrays-of-rotors-in-tpp-2014, bulk-inclusion-2016, doublerotors, sozzani-nmr, sozzani-xenon, Sozzani2004, surface-inclusion-2012, surface-inclusion-2013, surface-inclusion-2015, tpp-nanochannels, tpp-polymer, mythesis} (see \cref{fig:tpp-crystal}). The $a$ lattice parameter can expand up to roughly \SI{10}{\percent} to accommodate a variety of different sized guest molecules.\cite{arrays-of-rotors-in-tpp-2014,bulk-inclusion-2016} The lattice parameters also depend on the filling percentage of guests.\cite{doublerotors, mythesis} Recent efforts to grow TPP thin films requires rapid and frequent characterization of their structure and texture using GIXS on a tabletop system that utilizes a Cu K-$\alpha$ X-ray tube.\cite{mythesis}

\section{Scattering Enhancements}

\begin{figure}[b!]
    \centering
    \includegraphics{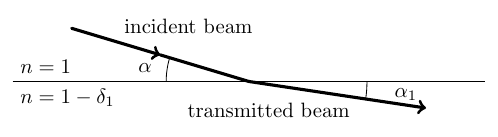}
    \caption[X-ray refraction in thin film]{
       Because the refractive index for X-rays traveling in materials is less than 1, the refraction angle $\alpha_1$ will always be smaller than the grazing-incidence angle $\alpha$ at void-material interfaces.
    }
\label{fig:snell}
\end{figure}

    Precise and accurate alignment of the film with respect to the beam is particularly necessary when performing GIXS on weak scatterers using laboratory-scale X-ray sources. Tilting the sample such that the beam takes a grazing angle of incidence $\alpha$ will enhance the interaction by projecting the beam over a large area across the sample. In the X-ray regime, materials have refractive indices less than unity, such that they can be described by
\begin{equation}
    n_i = 1 - \delta_i,
\end{equation}
    where $\delta_i\ll 1$ and $n_i$ is the index of refraction for material $i$.\cite{cullity3rd} Because the grazing-incidence angle $\alpha$ is the complimentary angle to the normal-incidence angle, Snell's law with respect to grazing angles is
\begin{equation}
    n_i\cos\alpha_i = n_j\cos\alpha_j.
\end{equation}
    For the first interface, between air and the top layer of a material stack (as seen in \cref{fig:snell}), Snell's law becomes
\begin{equation}
    \cos\alpha = (1-\delta_1)\cos\alpha_1.
\end{equation}
    At the critical angle, $\alpha_1=0$, so
\begin{equation}
    \cos\alpha_c = 1-\delta_1.
\end{equation}
    Because $\delta_1$ is very small, the small angle approximation can be used:
\begin{equation}
\label{eq:critical-angle}
    \alpha_c \approx \sqrt{2\delta_1}.
\end{equation}
    Therefore, at air-material interfaces, there will be a critical angle $\alpha_c$ for total external reflection.

    For maximum scattering intensity, the grazing-incidence angle should be greater than but very near the critical angle, such that the 
    transmitted beam travels near-parallel to the layer (\cref{fig:snell}) leading to the following enhancements to the scattering probabilities: (1) the beam being projected over the surface leads to an increased beam footprint, (2) the transmitted beam traveling near parallel to a long-axis of the material leads to long path-lengths, and (3) for sufficiently thin materials, reflections from the underside interface can lead to standing waves\cite{xray-standing-waves} to form due to the interference between the ``downward'' propagating beam and the ``upward'' propagating beam.
    


\section{Description of setup}

\begin{figure}[b!]
    \centering
    \includegraphics{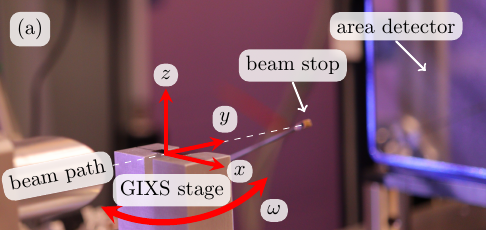}
    \includegraphics{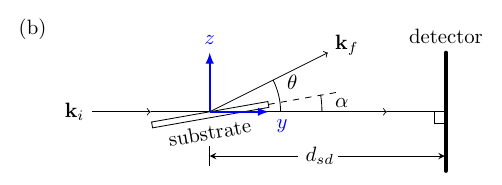}
    \caption{(a) The GIXS stage is controlled by $x$, $z$, and $\omega$ motors (as described in the text). The incident beam travels in the $y$-direction and is surface-normal to the area detector (in the $xz$-plane). The detector can move in the $x$, $y$, and $z$ directions. (b) The $\omega$ motor tilts the stage about an axis that is designed to coincide with the $x$-axis, thus setting the grazing-incidence angle $\alpha$. An exiting ray can be deflected to an angle $\theta$ due to reflecting from the surface or refracting when transmitting into the material.}
    \label{fig:setup}
\end{figure}

    Our GIXS system is shown in \cref{fig:setup}. A copper source is used to generate X-rays with a wavelength near $\lambda=\SI{1.54}{\angstrom}$. A Dectris EIGER R 1M detector with an array of 75 $\times$ \SI{75}{\micro\meter} pixels (covering an area of about \SI{62}{\square\centi\meter}) is used as an area detector and placed surface-normal to the incident beam (with incident X-rays propagating in the $y$-direction). There is a rectangular aperture that can be adjusted in both vertical and horizontal directions to set the beam size. For GIXS experiments, we set the aperture to \SI{100}{\micro\meter} in vertical width and 600-\SI{800}{\micro\meter} in horizontal width.
    
    A sample stage with vertical $z$, horizontal $x$, and angular $\omega$ (where $\omega$ is rotation about the $x$-axis) motor control is used to control the position and orientation of the sample relative to the beam. The $z$ motor is used to align the top surface of the sample with the beam, and the $\omega$ motor is used to set set the grazing-incidence angle. The detector can be moved in both the $x$ and $z$ directions to expand the field of view and the $y$ direction to set the sample detector distance $d_{sd}$.
    
    The system is controlled by \emph{spec}, a commonly used control software package for X-ray systems.\cite{spec} The beam stop contains a small area $(\sim\SI{1}{\square\milli\meter})$ PIN diode that is able to count a fraction of the primary incident X-rays to assist in the alignment procedure.

    An X-ray can be deflected upward at an angle $\theta$ due to a reflection or transmission event, and this deflected ray will hit the detector a distance $z$ above the beam center according to
\begin{equation}
    z = d_{sd} \tan\theta.
\end{equation}
    A specular reflection will occur at $z_s = d_{sd} \tan2\alpha$.


\section{\label{sec:scan}Typical scanning procedure}

\begin{figure}[b]
    \centering
    \includegraphics{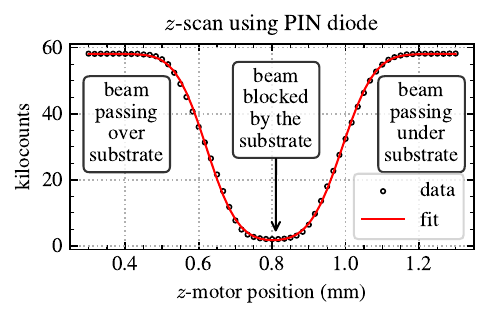}
    \caption{Example $z$-scan data using the PIN diode with a \SI{300}{\micro\meter} silicon substrate being moved through a beam with a vertical shutter set to \SI{100}{\micro\meter}. The sample can be aligned such that the top plane is at the center of the beam by placing the $z$-motor position such that the sample occludes half the total beam counts (approximately 29,000 counts in this scan and at 0.625 mm).}
    \label{fig:z-scan}
\end{figure}

    In the absence of specialty alignment equipment, a typical\cite{giwaxs-align} procedure for aligning the substrate to the incident X-ray beam would use the PIN diode or the area detector to detect the total X-ray flux passing the sample. It is often sufficient to align the $x$ motor visually because samples are larger than the beam in this direction. The \emph{spec} software package includes a program to scan through a motor position while measuring the X-ray counts on the beam stop's PIN diode (or an integrated area on the area detector) and plots counts versus motor position (see supplementary information). As seen in \cref{fig:z-scan}, scanning through $z$-motor positions shows that the sample occludes the beam as it passes through it. When the PIN diode sees half the total counts, the sample is occluding roughly half the beam, and therefore, the top surface will intersect the beam center at this corresponding $z$-motor position. Since this setup has the substrate span a gap, as the sample is brought up through the beam, it will eventually allow the beam to pass underneath it, so the counts will drop and then come back to the max value as the sample moves from below the beam to above the beam. This behavior can be modeled by two error functions. An error function, denoted erf$(x)$ is defined by
\begin{equation}
    \text{erf}(x) = \frac{2}{\sqrt{\pi}} \int_0^x e^{-t^2}dt.
\end{equation}
    The fitting function can be written
\begin{equation}
\begin{split}
\label{eq:I1}
    I_1(z) = \frac{I_\text{max} - I_\text{min}}{2}\bigg(&\text{erf}\bigg(\frac{z-z_\text{a}}{\sigma_a}\bigg) \\
    & - \text{erf}\bigg(\frac{z-z_\text{b}}{\sigma_b}\bigg)\bigg) + I_\text{max},
\end{split}
\end{equation}
    where $I_\text{max}$ is the average maximum counts when the beam is not occluded, $I_\text{min}$ is the average minimum counts when the beam is fully covered by the sample, $z_b$ is where the sample occludes the bottom half of the beam, $z_a$ is where the sample occludes the top half of the beam, and $\sigma_a$ and $\sigma_b$ represent the slope of each of the error functions.

    If the sample rests on a stage without a gap, such that the stage will occlude the beam when the sample is above the beam, then $I(z)$ can be modeled as a single error function:
\begin{equation}
\label{eq:I2}
    I_2(z) = -\frac{I_\text{max} - I_\text{min}}{2}\text{erf}\bigg(\frac{z-z_\text{a}}{\sigma_a}\bigg) + \frac{I_\text{max}+I_\text{min}}{2}.
\end{equation}

\begin{figure}[b!]
    \centering
    \includegraphics{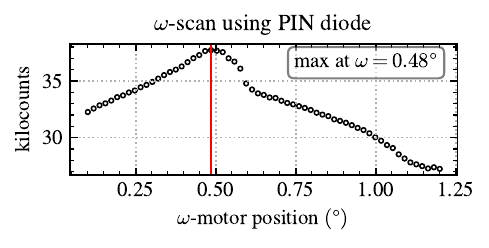}
    \caption{Example $\omega$-scan data using the PIN diode with a \SI{300}{\micro\meter} silicon substrate being rotated with an axis of rotation in the $x$-direction. The red line marks the motor angle corresponding to the maximum counts }
    \label{fig:om-scan}
\end{figure}

    The choice of fitting the $z$-scan with \cref{eq:I1} or \cref{eq:I2} is dependent on whether the sample stage will occlude the beam or not. Either way, the $z_a$ parameter in each fit will yield the $z$-motor position corresponding to the sample's upper surface being moved halfway into the beam.



\begin{figure}[t!]
    \centering
    \includegraphics{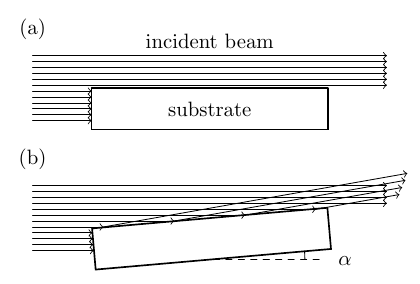}
    \caption{
        (a) When the substrate is parallel to the beam, half of the counts (represented as six rays) pass over the sample and half of the counts get occluded. (b) When tipped to a grazing-incidence angle below the critical angle, some of the rays that were passing over the substrate (two rays) hit the surface and reflect, such that they still reach the detector. However, there are also rays that were occluded at $\alpha=0$ that now hit the surface and reflect into the detector (two rays).
    }
\label{fig:refl-occl}
\end{figure}

    Once the sample is aligned in $x$ and $z$, an $\omega$-scan can be done to attempt to find the angle $\omega_0$ corresponding to the sample's top plane being parallel to the incident beam direction. An example dataset using the PIN diode to measure counts as a function of $\omega$ is plotted in \cref{fig:om-scan}. Based on the assumption that a parallel sample will occlude the beam by the least amount, the peak value in an $\omega$-scan should be $\omega_0$.\cite{giwaxs-align} However, for reflective materials, as the sample is tipped back, more cross-sectional area of the film's top plane will intersect the beam, resulting in more of the incident beam being specularly reflected (see \cref{fig:refl-occl}). This effect results in an increase in counts past $\omega_0$ until either the reflected beam starts to miss the PIN detector (resulting in less of the total counts being detected) or the critical angle for the substrate is reached (where some of the intensity that would have been reflected is now split between reflection and transmission followed by absorption).

\section{Spatially resolved scanning procedure}

    Alternative to measuring the total X-ray flux that passes the sample, as discussed above, we propose using the area detector to spatially resolve X-rays from (1) the direct beam, (2) the reflected beam, and (3) the transmitted beam. The spatially-resolved $z$-scan (SRS-$z$) or the spatially-resolved $\omega$-scan (SRS-$\omega$) are done by capturing images as a function of $z$-motor position or $\omega$-motor position in a similar way to the traditional $z$-scan and $\omega$-scan. For motor $m$ ($m$ being either $z$ or $\omega$), an SRS-$m$ is performed by running the following steps in a \emph{spec} macro (an example is shown in the supplementary information):

\begin{enumerate}
  \item Move the beamstop out of the path.
  \item Move $m$ to $m_i$ (starting motor position).
  \item\label{it:expose} Expose the detector for \SI{100}{\milli\second} and record the image along with the $m$-motor's current angle position.
  \item\label{it:incr} Increment $m$ by $\delta m$.
  \item Repeat steps \ref{it:expose} and \ref{it:incr} until $m>m_f$.
  \item Return $m$ to its original position.
  \item Move the sample out of the path.
  \item Expose the detector for \SI{100}{\milli\second} (this produces an image of the direct beam).
  \item Move the sample back to its aligned position.
  \item Return the beamstop to its original position.
\end{enumerate}

\begin{figure}[b!]
    \centering
    \includegraphics{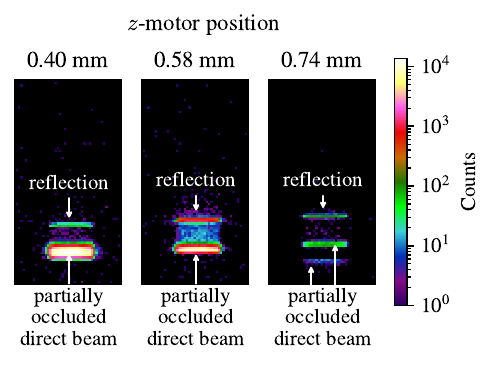}
    \caption{Example images ($6.4 \times \SI{3.3}{\milli\meter}$) from an SRS-$z$ with \SI{100}{\milli\second} exposure times (cropped using the method described in the supplementary information). As the sample moves upward, it occludes more of the beam. As the sample is raised upwards in the $z$-direction, the reflection shifts upward, and the brightness of the reflection will depend on the brightness of the beam that hits the top surface.}
    \label{fig:crop-z}
    
    \includegraphics{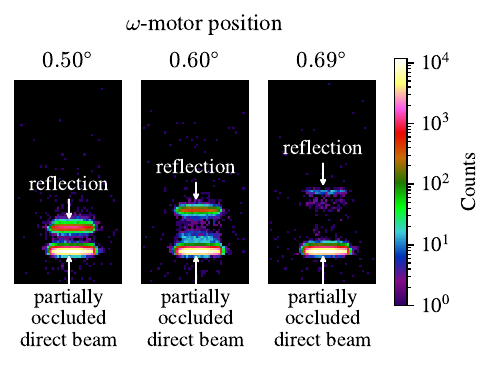}
    \caption{Example images ($6.4 \times \SI{3.3}{\milli\meter}$) from an SRS-$\omega$ with \SI{100}{\milli\second} exposure times (cropped using the method described in the supplementary information). The sample occludes the bottom half of the direct beam. As the stage is tipped further, the reflection is observed higher on the detector.}
    \label{fig:crop-om}
\end{figure}

    For a SRS-$z$, the images will show the sample occlude the beam as a function of motor position. If the sample is tilted such that there will be a reflection, the images will also show the intensity in the reflection correspond to the intensity of the beam that the top surface intersects. Examples are shown in \cref{fig:crop-z}. Similarly for an SRS-$\omega$, this procedure will result in a series of images where the reflection will hit the detector as a function of the incident angle, such as shown in \cref{fig:crop-om}.

\section{Alignment using spatially resolved scan}\label{sec:spec-align}

\begin{figure}[h!]
    \centering
    \includegraphics{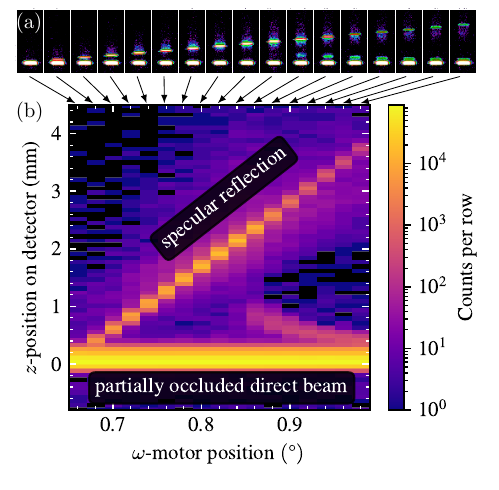}
    \caption{
        (a) Sequence of images exposed at increasing $\omega$-motor positions. (b) Individual images are summed row-wise and then concatenated column-wise to produce an array of intensities that are spatially resolved vertically as a function of $\omega$-motor position. The $z$-axis represents the vertical position on the detector. This is an SRS-$\omega$ on a silicon substrate with a \SI{300}{\nano\meter} thermal oxide layer.
    }
\label{fig:spec-plot-splainer}
\end{figure}

    \cref{fig:spec-plot-splainer}a shows an extended sequence of images as a function of the $\omega$-motor position for a silicon substrate with a thermal oxide layer. The counts in each row of each image are summed (producing a vertical array of counts per row for each image). The counts-per-row arrays are combined as columns to create a two-dimensional array where the column index represents the $\omega$-motor position and the row index represents vertical position on the detector (see \cref{fig:spec-plot-splainer}b).

    As with the alignment strategy outlined in \cref{sec:scan}, spatially-resolved alignment iterates between vertical (SRS-$z$) and angular (SRS-$\omega$) alignment scans until the alignment is sufficiently refined.

\subsection{Vertical alignment}

\begin{figure}[t!]
    \centering
    \includegraphics{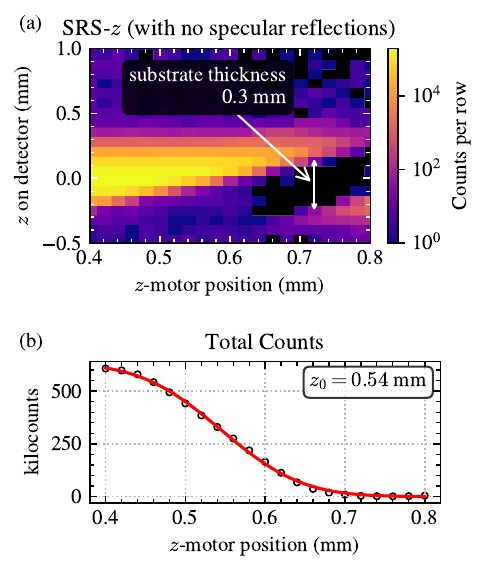}
    \caption{
        (a) When the $\omega$-motor position is set such that there is no observable specular reflection, the spatial information is not important, and the total counts versus $z$-motor position can be used to select the first vertical position for the alignment procedure. However, the thickness of the \SI{0.3}{\milli\meter} silicon substrate can be roughly observed in the image. (b) Adding the total counts from all the rows in each image of the scan reduces the data to typical $z$-scan.
    }
\label{fig:srs-z2}
\end{figure}

    To vertically align the sample, an SRS-$z$ is performed. If the sample is tipped to an angle such that no specular reflections will observably occur, as seen in \cref{fig:srs-z2}a, then the the spatial information simply shows the rising shadow from the substrate blocking the incident beam. In this case, the data can be reduced to the same $z$-scan discussed in \cref{sec:scan} by adding all the counts from each column (a sum over each of the entire images of the scan) (\cref{fig:srs-z2}b).

\begin{figure}[t!]
    \centering
    \includegraphics{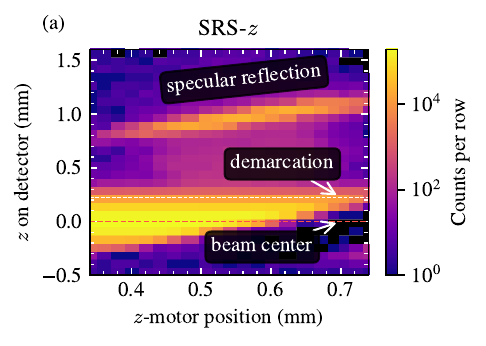}
    \includegraphics{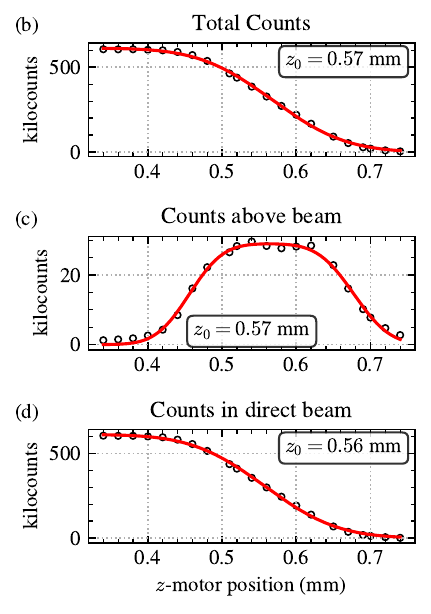}
    \caption{
        (a) An SRS-$z$ can be reduced to (b-d) three separate alignment plots to help determine the $z$-motor position corresponding to the reflecting surface intersecting the beam center. (b) The total counts give the result of a typical scan, discussed in \cref{sec:scan}. (c) The counts above the demarcation line is approximately the total rays reflected by the substrate. (d) The counts below the demarcation line better represents the assumptions made in \cref{sec:scan}. Small discrepancies can occur due to the scan not extending far enough in both directions.
    }
\label{fig:srs-z1}
\end{figure}
    
    However, if the sample is tipped to a positive grazing-incidence angle, specular reflections appear in the images. The SRS-$z$ data can then be reduced into three plots. An example of an SRS-$z$ on a silicon substrate with a \SI{300}{\nano\meter} thick thermal oxide layer with the three reduced plots are shown in \cref{fig:srs-z1}. The first reduced plot (\cref{fig:srs-z1}b) sums the total counts in each column and represents the data for a $z$-scan without spatially resolved data (as discussed in \cref{sec:scan}). A demarcation line is placed on the SRS-$z$ plot which separates ``above the beam'' and ``the direct beam.'' The demarcation line is one full vertical beam-width above the beam center (the determination of the beam center and width is shown in the supplementary information). The rows above the demarcation and below the demarcation are summed to produce the second two reduced plots (\cref{fig:srs-z1}c and \cref{fig:srs-z1}d respectively). In \cref{fig:srs-z2}c, the intensity of the specular reflection increases as the reflecting surface intersects brighter parts of the beam, and then the intensity falls back to zero as the reflecting surface moves above the brightest part of the incident beam. The total counts and the counts in the direct beam can be fitted with \cref{eq:I2}, and the counts above the beam can be fitted with
\begin{equation}
\label{eq:I3}
    I_\text{above}(z) = \frac{I_\text{max}}{2}\bigg(\text{erf}\bigg(\frac{z-z_\text{b}}{\sigma_b}\bigg) - \text{erf}\bigg(\frac{z-z_\text{a}}{\sigma_a}\bigg)\bigg),
\end{equation}
    where the reflecting surface can be considered to be centered in the beam at
\begin{equation}
    z_0=\frac{z_a+z_b}{2}.
\end{equation}


\subsection{Angular alignment}

\begin{figure}[b!]
    \centering
    \includegraphics{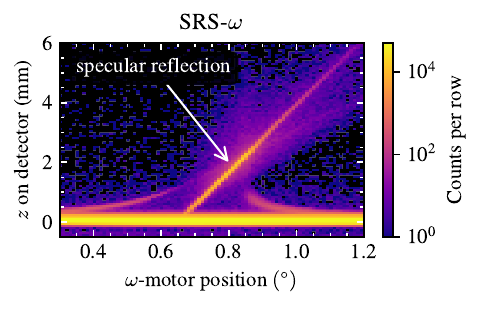}
    \caption{
        Example SRS-$\omega$ data for a silicon substrate with a \SI{300}{\nano\meter} thick thermal oxide layer. The figure shows the primary beam (yellow counts around $z=0$ on the detector), the specular reflection (rising diagonal counts starting at $z=0$ and $\omega\approx\SI{0.65}{\degree}$ and continuing to $z\approx\SI{6}{\milli\meter}$ and $\omega\approx\SI{1.2}{\degree}$), and other features to either side of the specular reflection that are described later in the text.
    }
\label{fig:srs-om1}
\end{figure}

    The SRS-$\omega$ is used to determine the motor position $\omega_0$ corresponding to the reflecting surface being parallel to the incident beam direction (i.e. when $\alpha=0$). The $\omega$-motor position can then be related to the grazing-incidence angle by
\begin{equation}
    \alpha = \omega - \omega_0.
\end{equation}
    An example SRS-$\omega$ result on a silicon substrate with a \SI{300}{\nano\meter} thick thermal oxide layer is shown in \cref{fig:srs-om1}. The figure shows the primary beam (yellow counts near $z=0$ on the detector), the counts due to the specular reflection (yellow counts along a diagonal), and counts that are deflected due to other effects like refraction as described in \cref{sec:transmission}. Here we concentrate on the specular reflection counts that can be seen rising linearly as a function of $\omega$. The trend of the reflection will follow
\begin{equation}
\label{eq:specular-fit}
    z = d_{sd} \tan{[2(\omega-\omega_0)]},
\end{equation}
    but at these very low angles, $\tan{x}\approx x$, so on the SRS-$\omega$ plot, the specular will follow a line of slope $2d_{sd}$ and will intersect the beam center ($z=0$) at $\omega_0$. Inspecting by eye shows that the trend intersects the beam center somewhere between \SI{0.6}{\degree} and \SI{0.7}{\degree}.

    The total counts that pass the sample with respect to the $\omega$-motor angle is often quoted as having a maximum when the sample is flat.\cite{giwaxs-align} This assumes there are no reflections from the sample and that half the beam will pass over the substrate when it is flat with respect to the beam ($\alpha=0$), and when it is tipped in either direction, the back or front corner will protrude up into the beam and occlude more of it. However, as seen in \cref{fig:refl-occl,fig:srs-om1}, as the substrate tips to $\alpha>0$, rays that were previously occluded hit the top surface and are reflected into the detector. Therefore, as the substrate is tipped above $\alpha_0$, the counts should continue to increase. It is only above the critical angle, when X-rays begin to transmit into the material, that X-rays that hit the top surface will begin to be absorbed significantly and the counts will begin to decrease.

\begin{figure}[b!]
    \centering
    \includegraphics{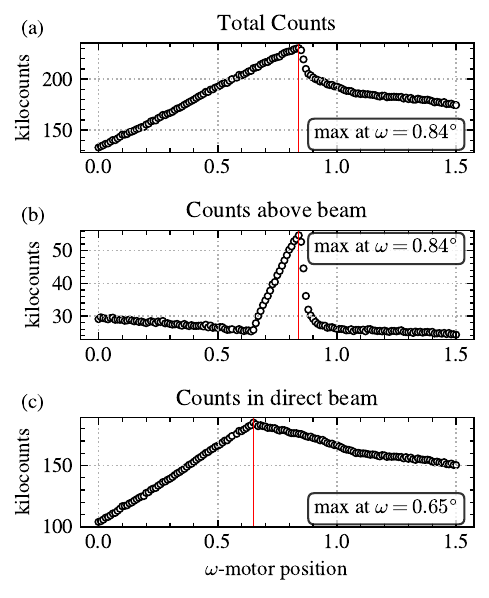}
    \caption{
        The SRS-$\omega$ data reduced by (a) summing the total counts of each exposure, (b) summing the counts above the demarcation line (one vertical beam width above the beam center), and (c) summing the counts below the demarcation line. The max value above the beam should occur at an angle just below the critical angle, and the max value below the beam should occur approximately at $\omega_0$.
    }
\label{fig:srs-om-redu}
\end{figure}

\begin{figure}[b!]
    \centering
    \includegraphics{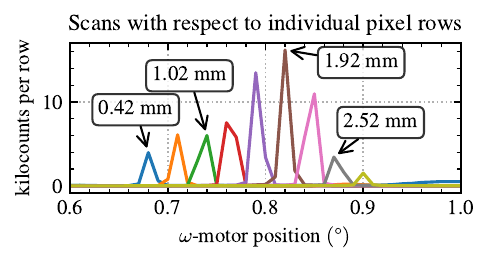}
    \caption{
        Each row of the SRS-$\omega$ data represents the X-ray intensity at a certain vertical position on the detector as a function of the motor position. This graph shows a handful of these rows plotted versus $\omega$-motor position. Each of these curves shows the $\omega$-motor position corresponding to the specular reflection occurring at that vertical position. 
    }
\label{fig:srs-om-lines}
\end{figure}

\begin{figure}[b!]
    \centering
    \includegraphics{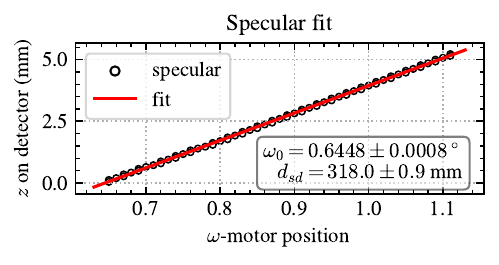}
    \caption{
        For each $z$ position on the detector, the $\omega$-motor position corresponding to the maximum counts is selected as a data point. \cref{eq:specular-fit} is fitted to the data for $\omega_0$ and $d_{sd}$, listed with uncertainties. These results are consistent with the qualitative results for $\omega_0$ and $d_{sd}$ as described in the text and shown in \cref{fig:srs-om-redu}.
    }
\label{fig:spec-fit1}
\end{figure}

\begin{figure*}
    \centering
    \includegraphics{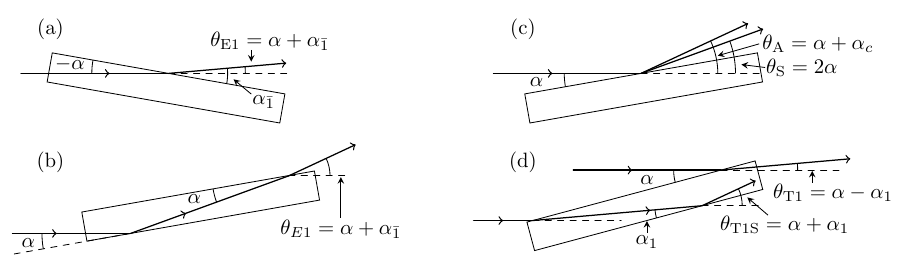}
    \caption{
        (a) When the grazing-incidence angle is negative, X-rays that enter through the side wall can be refracted while exiting the material (deflected by $\theta_\text{E1}$). (b) At positive grazing-incidence angles, X-rays can enter through the side wall, specularly reflect from the bottom surface, and then refract out the top surface (deflected by $\theta_\text{E1}$). (c) At positive grazing-incidence angles, there will also be specular reflections (deflected by $\theta_\text{S}$), and there can be anomalous surface reflections (deflected by $\theta_\text{A}$). (d) When the grazing-incidence angle exceeds the critical angle, along with the specular and anomalous reflections, there will be transmission into the material. This leads to two exiting angles: (1) where the beam refracts and then exits the sample (deflected by $\theta_\text{T1}$) and (2) where the beam refracts, reflects off the bottom surface, and then exits the sample (deflected by $\theta_\text{T1S}$).
    }
\label{fig:reflection-geo}
\end{figure*}
    
    By parsing the total counts above and below the demarcation line (placed one vertical beam-width above the beam center), as seen in \cref{fig:srs-om-redu}, it is clear that the counts in the direct beam and the total counts peak at different angles. The counts below the demarcation line should mostly ignore the reflected X-rays and peak very near $\omega_0$. In the case of a single layer of surface localized material (as we have in this example, with the thermal oxide thin film), then the counts above the beam should peak just under the critical angle for that material because the peak value will correspond to the sample being tipped to the highest grazing-incidence angle before transmission occurs. Therefore, the counts below the demarcation line can be used to approximate $\omega_0$ by finding the motor position corresponding to the maximum counts, and for a single layer, the counts above the beam can be used to approximate the critical angle. For the SRS-$\omega$ data shown in \cref{fig:srs-om1}, the max values (from \cref{fig:srs-om-redu}) suggest that $\omega_0\approx\SI{0.65}{\degree}$ and $\alpha_c\approx\SI{0.19}{\degree}$.

    The specular reflection position on the image can be fitted with \cref{eq:specular-fit} to extract $\omega_0$ and $d_{sd}$. This can most easily be done by locating the $\omega$-motor position corresponding to the maximum counts for each $z$ position on the detector. As seen in \cref{fig:srs-om-lines}, where individual rows of \cref{fig:srs-om1} are plotted, the peak value in each row corresponds to the specular reflection. The $\omega$-motor position corresponding to the maximum counts for each $z$ position on the detector is plotted in \cref{fig:spec-fit1} and are fitted using \cref{eq:specular-fit} for $\omega_0$ and $d_{sd}$.

    In this way, the specular reflection in the SRS-$\omega$ can be located and fitted to calibrate a GIXS system by determining the zero-angle and the sample-detector distance.

\section{Beam deflections due to additional processes}\label{sec:transmission}

    Along with the specular reflection, X-rays can also deflect from the incident beam's path due to refraction or anomalous surface reflection, both of which are clearly visible in \cref{fig:srs-om1}. \cref{fig:reflection-geo} shows these additional possible beam deflection paths due to a topmost layer (such as the thermal oxide thin film discussed in \cref{sec:spec-align}). When the sample is tipped forward (\cref{fig:reflection-geo}a), such that $\alpha \le 0$, the incident beam can enter the material through the side wall; Snell's law predicts that at near normal incidence, the beam won't deflect appreciably, so it continues in the same direction. Therefore, the beam will cross the material-air interface at the top surface with an incidence angle of $-\alpha$ (where $\alpha$ is negative), so
\begin{equation}
    (1-\delta_1)\cos\alpha=\cos\alpha_{\bar{1}}.
\end{equation}
    where $n_1=1-\delta_1$ is the index of refraction for the material. \cref{eq:critical-angle} can be used to relate $\delta_1$ to the critical angle. The small angle approximation leads to
\begin{equation}
    \label{eq:exiting-refraction}
    \alpha_{\bar{1}} = \sqrt{\alpha^2(1-\delta_1)+2\delta_{1}}.
\end{equation}
    The angle $\theta_\text{E1}$ above the incident beam that the refracted beam will travel will be
\begin{equation}
\label{eq:exiting1}
    \theta_\text{E1} = \alpha + \alpha_{\bar{1}} = \alpha + \sqrt{\alpha^2(1-\delta_1)+2\delta_1},
\end{equation}
    where $\alpha$ is negatively valued.

    At positive $\alpha$, the incident X-rays still pass into the side wall of the material and then reflect from the material-substrate interface (see \cref{fig:reflection-geo}b). The reflected beam will meet the material-air interface at an angle $\alpha$ and refract according to \cref{eq:exiting-refraction}. This will lead to a deflection also described by \cref{eq:exiting1} (but now with positive $\alpha$).

    X-rays that strike the top surface of the material can reflect via specular and anomalous surface reflection\cite{Yoneda1963} processes (as seen in \cref{fig:reflection-geo}c). Specular reflections will travel an angle of $2\alpha$ above the incident beam:
\begin{equation}
\label{eq:specular}
    \theta_\text{S} = 2\alpha.
\end{equation}
    Anomalous surface reflections occur at grazing-incidence angles and leave the surface at an angle equal to the critical angle for the material being reflected from, so the angle $\theta_\text{A}$ that the anomalous surface reflection will be deflected is
\begin{equation}
\label{eq:yoneda}
    \theta_\text{A} = \alpha + \alpha_{c}.
\end{equation}

    For grazing-incidence angles above the critical angle for the material ($\alpha\ge\alpha_c$), part of the beam will also penetrate into the sample (as seen in \cref{fig:reflection-geo}d). Snell's law predicts a refraction angle of
\begin{equation}
    \alpha_1 = \sqrt{\frac{\alpha^2-2\delta_1}{1-\delta_1}}.
\end{equation}
    The refracted beam can either leave the side wall, or further reflect off the material-substrate interface before leaving the side wall. This leads to the transmitted beam splitting into two paths
\begin{align}
\label{eq:transmission1}
    \theta_\text{T1} &= \alpha - \sqrt{\frac{\alpha^2-2\delta_1}{1-\delta_1}}\\
\label{eq:transmission-reflection1}
    \theta_\text{T1S} &= \alpha + \sqrt{\frac{\alpha^2-2\delta_1}{1-\delta_1}}.
\end{align}

\begin{figure}[h!]
    \centering
    \includegraphics{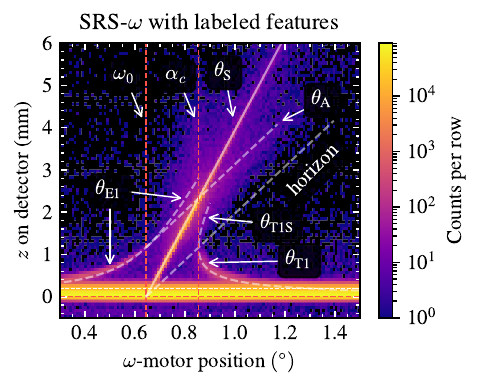}
    \caption{
        The same SRS-$\omega$ plot as \cref{fig:srs-om1} with labeled features, using $\alpha_c=\SI{0.21}{\degree}$. The dashed line labeled ``horizon'' marks a the projection tangent to the substrate surface toward the detector.
    }
\label{fig:srs-om-labeled}
\end{figure}

    \cref{fig:srs-om-labeled} shows the same SRS-$\omega$ as \cref{fig:srs-om1} of a silicon substrate with a \SI{300}{\nano\meter} thick thermal oxide layer but with all of the deflections labeled using $\delta_1 = 6.7\times 10^{-6}$ (corresponding to $\alpha_c=\SI{.21}{\degree}$).

\section{Deflections due to additional layers}

\begin{figure}[b!]
    \centering
    \includegraphics{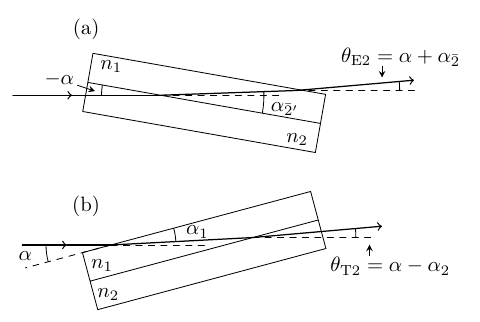}
    \caption{
        (a) An X-ray that refracts up from layer-2 into layer-1 and then out into the air will refact as if it was refracted due to a layer-2/air interface. (b) An X-ray that refracts down into layer-1 and then into layer-2 will be deflected as if it were refracted at an air/layer-2 interface.
    }
\label{fig:2layer}
\end{figure}

    We grow TPP thin films on the existing thermal oxide thin film, creating a two-layer stack. For a two-layer stack, the top layer (layer-1) will continue to result in all the deflections discussed above. Deflections due to the bottom layer (layer-2) are shown in \cref{fig:2layer}. The first layer has an index of refraction of $n_1=1-\delta_1$ and the second layer has an index of refraction of $n_2=1-\delta_2$.

    At negative grazing-incidence angles (\cref{fig:2layer}a) the incident beam can enter into the sides of both layers result in refracted beams that deviate slightly above the specular reflection. X-rays that enter the side of layer-1 will refract according to \cref{eq:exiting1}, but X-rays that enter the side of layer-2 will first refract at the layer-1/layer-2 interface and then at the layer-1/air interface. Snell's law then predicts
\begin{align}
    (1-\delta_2) \cos\alpha &= (1-\delta_1)\cos\alpha_{\bar{2}'}\\
    (1-\delta_1)\cos\alpha_{\bar{2}'} &= \cos\alpha_{\bar{2}}.
\end{align}
    Since $(1-\delta_2) \cos\alpha=\cos\alpha_{\bar{2}}$,
\begin{equation}
    \alpha_{\bar{2}} = \sqrt{\alpha(1-\delta_2) + 2\delta_1}.
\end{equation}
    Therefore, for an X-ray going into the side wall of layer-$i$ and then exiting the top surface of layer-1, the exiting angle will be
\begin{equation}
    \theta_{\text{E}i} = \alpha + \sqrt{\alpha^2(1-\delta_i)+2\delta_i}.
\end{equation}

\begin{figure}[b!]
    \centering
    \includegraphics{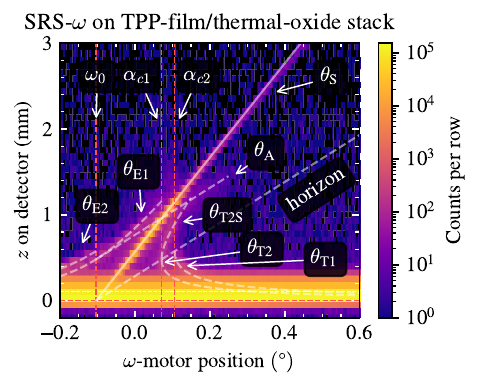}
    \caption{
        A two-layer material stack will show transmission curves for each of the two materials. However, $\theta_\text{E2}$ and $\theta_\text{T2S}$ do not appear because most of the X-rays are absorbed or scattered before reaching the end of the material when going through both layers. The curves are placed using $\delta_2=4.4\times 10^{-6}$ ($\alpha_{c1}=\SI{0.17}{\degree}$) for TPP and $\delta_2=6.7\times 10^{-6}$ ($\alpha_{c1}=\SI{0.21}{\degree}$) for the oxide layer beneath it.
    }
\label{fig:srs-om2}
\end{figure}

\begin{figure*}
    \centering
    \includegraphics{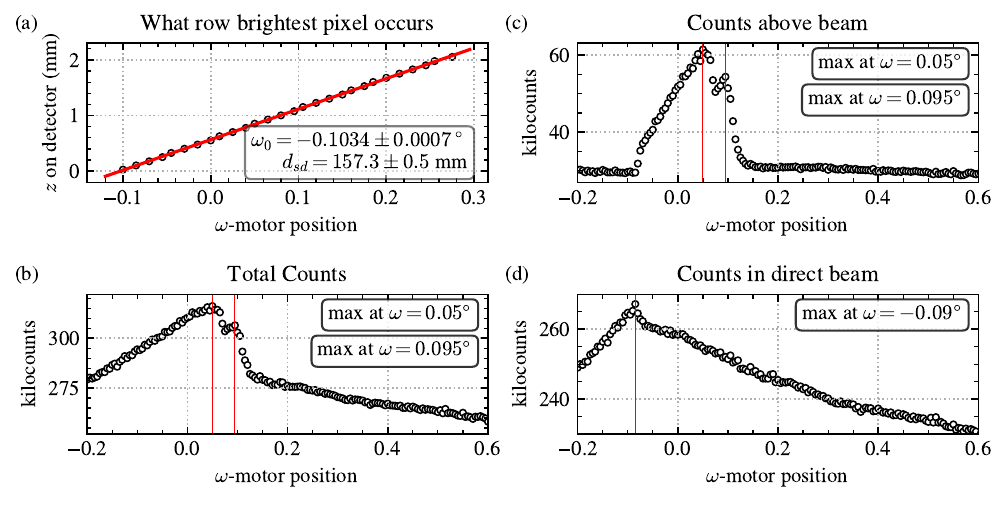}
    \caption{
        (a) Fit of the specular beam from the SRS-$\omega$ shown in \cref{fig:srs-om2}. (b) The total counts represents a typical scan and shows a max value over a tenth of degree above $\omega_0$. (c) Above the demarcation line, the peaks occur at the same angles. (d) The counts below the demarcation line more closely matches the assumptions made in \cref{sec:scan} fit of the specular beam and the  in the direct beam disagree by about \SI{0.013}{\degree}. This discrepancy is due to the the sample not being well centered, vertically, in the beam, and a further SRS-$z$ refinement step is needed. 
    }
\label{fig:srs-om2-redu}
\end{figure*}

    An X-ray that refracts at the air/layer-1 interface will refract according to
\begin{equation}
    \cos\alpha = (1-\delta_1)\cos\alpha_1
\end{equation}
    and can continue on to refract at the layer-1/layer-2 interface according to
\begin{equation}
    (1-\delta_1)\cos\alpha_1 = (1-\delta_2)\cos\alpha_2,
\end{equation}
    so
\begin{equation}
    \alpha_2 = \sqrt{\frac{\alpha^2-2\delta_1}{1-\delta_2}}.
\end{equation}

    An SRS-$\omega$ on a TPP/silica stack is shown in \cref{fig:srs-om2} with the features labeled and \cref{fig:srs-om2-redu} shows the fitting result of the specular beam as well as the reduced scans above and below the demarcation line.


\section{Impacts of accurate alignment}\label{sec:impact}

    Even without the $\omega_0$ determination, the SRS-$\omega$ can be used to determine the $\omega$-motor position corresponding to very near the critical angle. For low-$Z$ materials, like TPP, with low X-ray absorption, the transmission curve will bend vertically at the critical angle as the transmitted beam splits into $\theta_\text{T1}$ and $\theta_\text{T1S}$ (as seen in \cref{fig:srs-om2}). By locating these features, a user can select an $\omega$-motor position very near the critical angle of the relevant material layer in order to optimize scattering results.

\begin{figure}[h!]
    \centering
    \includegraphics{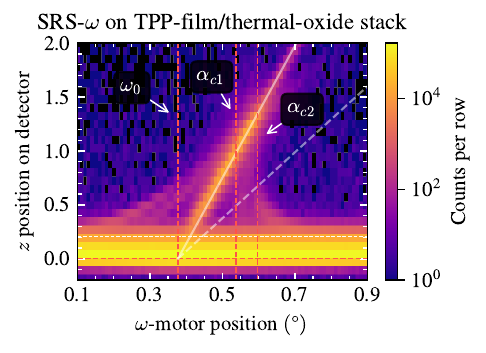}
    \caption{
        This SRS-$\omega$ scan was performed on a TPP film grown on a silicon substrate with a \SI{300}{\nano\meter} thick thermal oxide layer. The critical angles were determined by locating where the transmission curves meet the horizon.
    }
\label{fig:example-srs-om-for-compare}
\end{figure}

\begin{figure}
    \centering
    \includegraphics{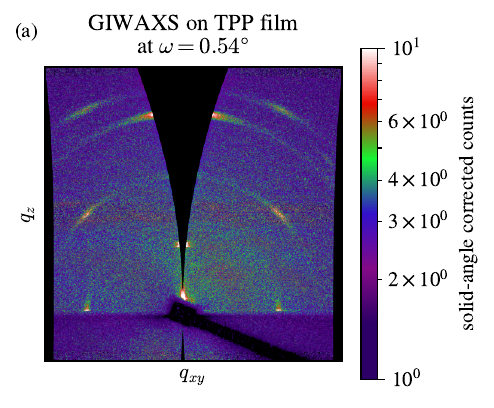}
    \includegraphics{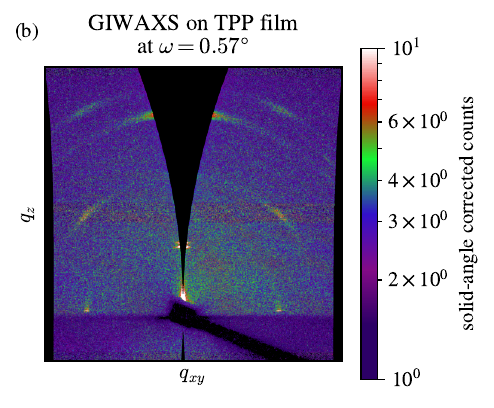}
    \caption{
        Example GIWAXS exposures (20 minutes) of the TPP film that was scanned in \cref{fig:example-srs-om-for-compare} at both (a) $\omega=\SI{0.54}{\degree}$ very near the critical angle and (b) a few hundredths of a degree above where the brightness in the scattering is reduced. Missing-wedge transformations convert the axes to reciprocal space (with a solid-angle correction applied\cite{mythesis}).
    }
\label{fig:giwaxs-eg}
\end{figure}

    \cref{fig:example-srs-om-for-compare} shows another SRS-$\omega$ plot from a TPP-film/thermal-oxide stack with user selected critical angles for the TPP ($\alpha_{c1}$) and the thermal oxide ($\alpha_{c2}$) that were determined by identifying $\theta_\text{T1}$, $\theta_\text{T1S}$, and $\theta_\text{T2}$. After the SRS-$\omega$ scan, eight GIXS exposures were performed, each at a different $\omega$-motor position, ranging from \SI{0.53}{\degree} to \SI{0.60}{\degree}. Two of these resulting images are shown in \cref{fig:giwaxs-eg} with missing-wedge transformations applied.\cite{how-to-giwaxs,film-texture-sim}

\begin{figure*}[t!]
    \centering
    \includegraphics{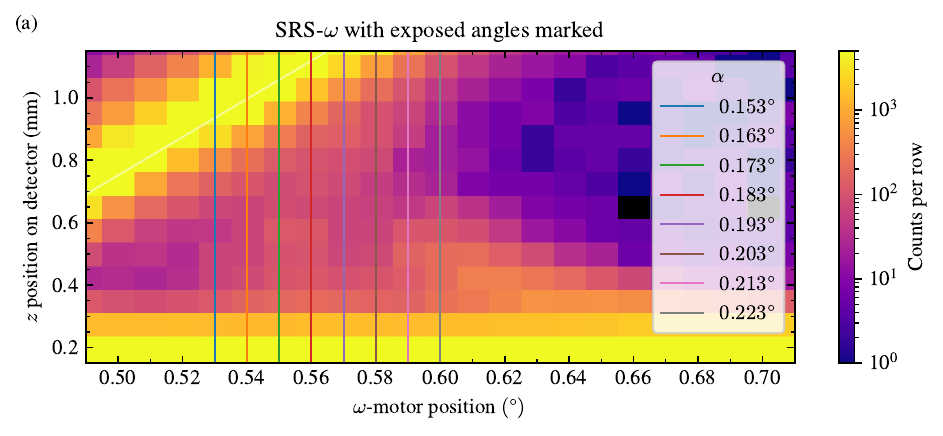}
    \includegraphics{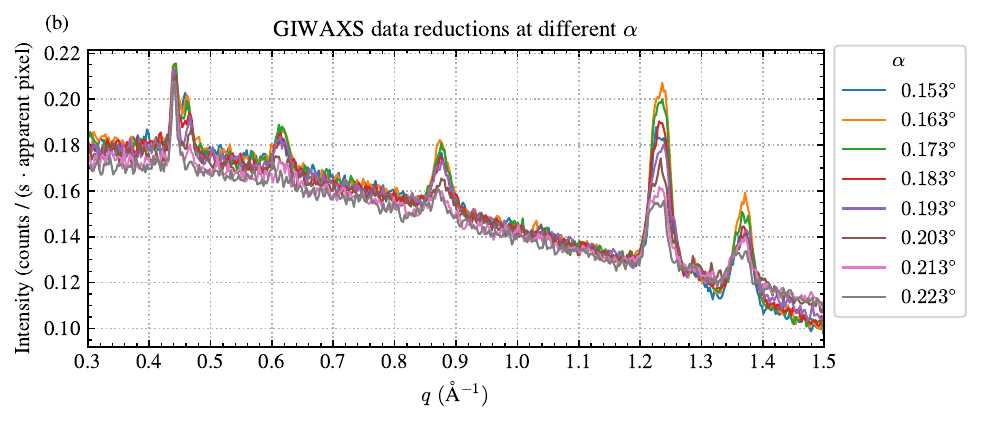}
    \caption{
        (a) The same SRS-$\omega$ scan shown in \cref{fig:example-srs-om-for-compare} zoomed in around the refraction features. The vertical lines mark the angles that were chosen to expose for GIWAXS (20 minutes each). Grazing-incidence angles are determined by subtracting $\omega_0$ from the SRS-$\omega$ fit from the $\omega$-motor positions. (b) Azimuthal integrations of each of the eight GIWAXS exposures, taken at the different grazing-incidence angles. The total intensity in the scattered peaks is sensitive to the grazing-incidence angle and is optimized near the critical angle of the TPP (roughly \SI{0.16}{\degree}).
    }
\label{fig:red-vs-om}
\end{figure*}

    Each of the eight GIXS exposures was reduced via azimuthal integration using pyFAI.\cite{pyFAI} These results are plotted together in \cref{fig:red-vs-om}, which shows the sensitivity of GIXS experiments to the choice of $\alpha$. Choosing the $\omega$-motor position corresponding to when transmission clearly begins to appear in the SRS-$\omega$ results in the most scattering intensity, and the scattering intensity is sensitive to \SI{0.01}{\degree} shifts in incident angle. A shift of \SI{0.05}{\degree} corresponds to roughly half the counts in a peak relative to the background.

\section{Conclusion}

    The results in \cref{sec:impact} indicate how sensitive the scattering intensity is to the alignment of the grazing-incidence angle, where the maximal result occurs very near the critical angle. When using weak X-ray sources on materials with low electron densities, it is particularly important to optimize the scattering intensity through careful and precise alignment of the incident angle.
    SRS-$\omega$ as described here is easily done use the existing area detector and
    not only allows a user to calibrate the system so that $\omega_0$ and $d_{sd}$ are known, but it also simultaneously gives information on the critical angle for the material and what motor position will be best suited for maximizing the scattering intensity.

\twocolumngrid 
\pagebreak
\appendix
\pagebreak
\onecolumngrid
\pagebreak

\section{Supplementary information}

\subsection{Finding the beam center}
\label{sec:crop}

    Locating the center of the beam on the detector allows the images to be cropped around the beam, such that only relevant data is preserved. First the counts per column and counts per row are fitted, each with a Gaussian function, to determine roughly where the beam peaks in both vertical and horizontal directions on the image. The location can be refined by using the following fitting function:
\begin{equation}
\label{eq:beam-finder}
    \frac{A}{2}\bigg[\text{erf}\bigg(\frac{x - x_0 + w/2}{\sigma}\bigg)-\text{erf}\bigg(\frac{x - x_0 - w/2}{\sigma}\bigg)\bigg],
\end{equation}
    where $A$ is the amplitude, $x_0$ is the beam center, $w$ is the beam width.

\subsection{Cropping images}

    The amount of appropriate cropping relative to the beam center can be determined by the sample-detector distance $d_{sd}$, the intended scanning range of $\omega$-motor positions, and the pixel size (with horizontal width $w_p$ and vertical height $h_p$). At the maximum angle in the scan, the specular reflection will hit the screen at a distance above the sample according to the vertical distance the beam travels ($d_{sd}\tan{2\alpha}$), so an estimate of the range of rows above the beam center where observable reflections will occur can be calculated using:
\begin{equation}
    \text{rows}_\text{above} > \frac{d_{sd}}{h_p}\tan2\alpha_m,
\end{equation}
    where $d_{sd}$ should be approximately known, and $\alpha_m$ is an estimated maximum grazing-incidence angle during the scan (one can often safely assume $\alpha_m=\SI{1}{\degree}$). Only a few pixels below the beam need to be preserved. To be safe, this can be estimated by five times the vertical width of the beam $w_z$ (as determined by \cref{eq:beam-finder}):
\begin{equation}
    \text{rows}_\text{below} = 5w_z.
\end{equation}
    The number of columns to keep can be $4w_s/w_p$ centered around the beam center, where $w_s$ is the horizontal width of the shutter.

\begin{figure}[h!]
    \centering
    \includegraphics{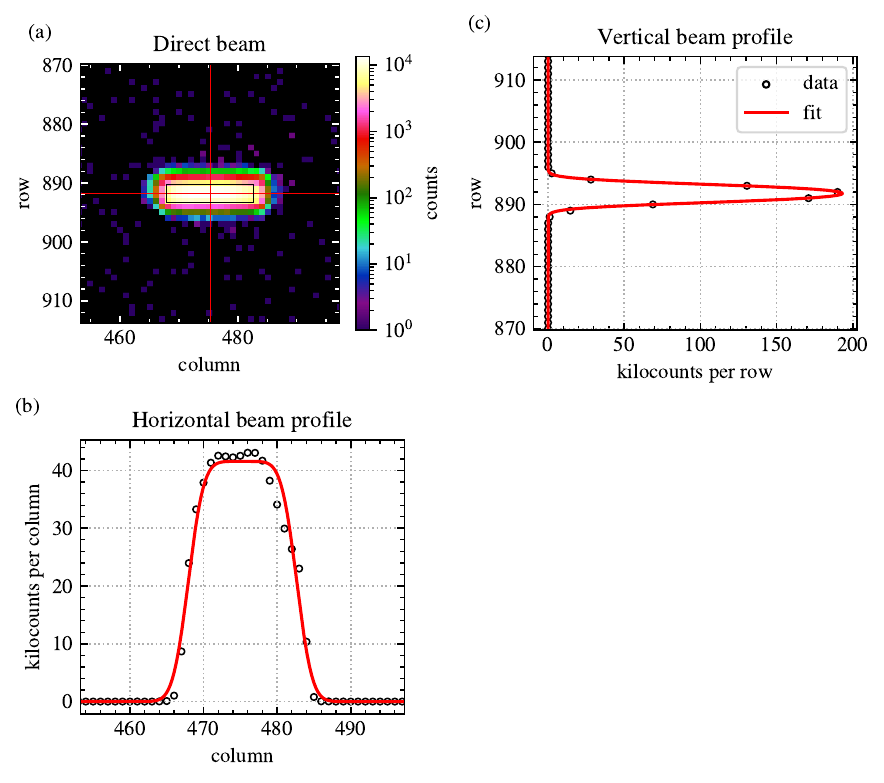}
    \caption{(a) The Cu-K$\alpha$ X-ray beam imaged directly on the detector (with $75 \times \SI{75}{\micro\meter}$ pixels). The beam is located by fitting \cref{eq:beam-finder} through (b) the horizontal and (c) vertical beam profiles. The red cross in the image shows the fitted center, and the black rectangle inside the beam shows the fitted beam width in both directions (based on the center of the erf functions described in the text).}
    \label{fig:beam-center}
\end{figure}

\subsection{scan macro example}

    The following is an example macro that would be saved as a text file. It can be called in \emph{spec} with ``\codeword{do path/to/macro.txt}''.

\begin{verbatim}
    umv s2vg 0.1
    umv s2hg 0.6
    umv wbs 5.0000
    umv om -0.4000
    eiger_run 0.1 om_scan_-0_4_degrees.tif
    umv om 0.0000
    eiger_run 0.1 om_scan_0_0_degrees.tif
    umv om 0.4000
    eiger_run 0.1 om_scan_0_4_degrees.tif
    umv om 0.8000
    eiger_run 0.1 om_scan_0_8_degrees.tif
    umv om 1.2000
    eiger_run 0.1 om_scan_1_2_degrees.tif
    umvr z -10.0000
    eiger_run 0.1 om_scan_direct_beam.tif
    umvr z 10.0000
    umv wbs -5.0000
    umv om -4.0000
\end{verbatim}

    First the shutters are set to \SI{0.1}{\milli\meter} vertically and \SI{0.6}{\milli\meter} horizontally and the beam stop is moved out of the way. Then $\omega$-motor positions are set to values between \SI{-0.4}{\degree} and \SI{1.2}{\degree} with \SI{0.4}{\degree} steps. In this case, \codeword{eiger_run [exposure time] [filename]} is a program to expose the detector, and the $\omega$-motor position is encoded in the filename. In some implementations, it is possible to save the images as EDF files, where the motor position can be stated in the file's header. Finally, the sample is moved down out of the beam, so that the the direct beam can be imaged, and then the sample is returned, the beam stop is returned, and the motor position is moved back to the start of the scan.

    The motor position can be parsed from the filename using the following Python function
\begin{lstlisting}[language=Python]
    def motor_angle_from_filename(filename: str):
        left_of_decimal = filename.split("_")[-3]
        angle = float(left_of_decimal)
        right_of_decimal = (filename
                            .split("_")[-2].replace(".tif", ""))
        if left_of_decimal[0] == "-":
            angle -= (float(right_of_decimal) 
                      / 10. ** len(right_of_decimal))
        else:
            angle += (float(right_of_decimal) 
                      / 10. ** len(right_of_decimal))
        angle = round(angle, 3)
        return angle
\end{lstlisting}
    The motor position can be pulled from the comment of an EDF file's header with the assistance of the FabIO\cite{fabio} and the re\cite{re} Python packages using the following function

\begin{lstlisting}[language=Python]
    import fabio
    import re

    def motor_angle_from_file_header(filename: str):
        comment = fabio.open(filename).header["Comment"]
        angle = float(
            re.search(r"[-+]?\d*\.\d+|[-+]?\d+", comment).group()
        )
        return angle
\end{lstlisting}
    This function will pull a decimal number (and recognize a minus sign) from the comment in the header.

\bibliographystyle{aipnum4-1}
\bibliography{bib}

\end{document}